\documentclass[11pt]{article}
\usepackage{moriond,epsfig}

\def\Journal#1#2#3#4{{\em #1} {\bf #2}, #3 (#4)}

\def\micron{$\mu$m }
\def\pn{\par\noindent}
\def\etal{\textit{et al.}}

\def\be{\begin{equation}}
\def\ee{\end{equation}}
\def\bea{\begin{eqnarray}}
\def\eea{\end{eqnarray}}

\def\NH{N$_{H}$ }
\def\due{$^{-2}$}
\def\uno{$^{-1}$}

\begin{document}
\vspace*{4cm}
\title{THE CONTRIBUTION OF ACTIVE GALAXIES TO THE FAR--INFRARED 
AND SUBMILLIMETER BACKGROUNDS}

\author{ M. BRUSA$^{1}$, A. COMASTRI$^{2}$, C. VIGNALI$^{3}$}

\address{$^{1}$ Dipartimento di Astronomia, Universit\`a di Bologna, via Ranzani 1,
40127 Bologna, Italy}
\address{$^{2}$ Osservatorio Astronomico di Bologna, via Ranzani 1,
40127 Bologna, Italy}
\address{$^{3}$ Department of Astronomy and Astrophysics, The Pennsylvania State 
University, 525 Davey Lab, University Park, PA 16802, USA}

\maketitle\abstracts{
We present a simple model, developed to estimate the contribution
of Active Galactic Nuclei to the far--infrared background,
closely linked to the AGN synthesis model for the X--ray background.
According to our calculation the AGN contribution is never dominant,
ranging from a few to 15 percent between 100 and 850 $\mu$m.}

\section{Introduction}

Deep X--ray surveys carried out with \textit{Chandra} and
XMM--\textit{Newton} have resolved into discrete sources
a large fraction (more than 75\%) of the hard 2--10 keV X--ray
background (XRB) (Mushotzky \etal~2000, Giacconi \etal~2001, 
Hasinger \etal~2001).\\
Most of the so far optically identified objects are Active
Galactic Nuclei (AGN) and a sizeble fraction of them
shows indication of nuclear obscuration, in relatively good agreement
with the expectation of AGN synthesis models of the XRB (Setti \& Woltjer~1989,
Madau, Ghisellini \& Fabian~1994, Comastri \etal~1995, Gilli \etal~2001).\\
The nuclear radiation of the obscured AGN responsible for the majority of the
XRB spectral intensity must be re--irradiated by dust in the far--Infrared
(FIR) and submillimeter (submm) bands.
It is thus likely that absorbed AGN provide a significant
contribution to the Cosmic Far Infrared Background (CFIRB)
recently measured by COBE between 100 and 2000 \micron (Puget \etal~1996, 
Fixsen \etal~1998, Hughes \etal~1998, Lagache \etal~2000) and to the 
850 \micron SCUBA source counts (Smail, Ivison \& Blain~1997).\\
In order to quantitatively estimate the contribution of 
X--ray (obscured) sources to the CFIRB, a simple synthesis
model has been developed. 
Our strategy is to link the FIR and X--ray spectral
properties of a representative sample of AGN
and then evaluate their contribution to the CFIRB adopting 
the same assumptions used by Comastri \etal~(1995) to fit the XRB.
More specifically, since a key parameter in this model is
the distribution of objects as a function of their X--ray 
column density (\NH, in the range $10^{21}-10^{25}$ cm\due), the 
CFIRB model has been computed keeping the absorption distribution
which provides the best fit to the XRB observational constraints 
and looking for correlations between the FIR and X--ray properties 
for each class of \NH.

\section{The model}

 The intensity of diffuse emission due to discrete sources 
 at an observed frequency $\nu_0$ 
 can be written as:

\be
 I(\nu_0)=\frac{c}{4\pi H_0} \int_0^{z_{max}} 
	\int_{L_{min}}^{L_{max}} \frac{dV}{dz} L \rho(L,z)
        g(\nu)dLdz
 \label{eq:fondo}
 \ee
\pn
 where $g(\nu=\nu_0(1+z))$ is the source spectral shape 
 and $\rho(L,z)$ describes the luminosity function and its
 evolution with the redshift.
\pn
 The FIR spectrum is modelled with a single--temperature,
 optically--thin greybody curve ($g(\nu)\propto B_{\nu}(T)\nu^{\beta}$,
 where $B_{\nu}(T)$ is the Planck function),
 which is appropriate to describe thermal re--radiation of primary 
 emission from dust.
 Fiducial values for the dust temperature T and the emissivity 
 index $\beta$ were derived by fitting the FIR (longward of 60 $\mu$m)
 and 850 \micron literature data of a large sample of about 100 
 nearby, hard X--ray selected AGN, almost equally
 populating the four classes of absorption column density used 
 in the XRB model (centered at $\log{N_H}=21.5, 22.5, 23.5, 24.5$).
 The results indicate a narrow range in the best--fit values for
 both the dust temperature (T=30--50 K) and the emissivity index
 $\beta$ (1--2);
 moreover these values are independent from X--ray absorption
 and luminosity. The median values (T=40 K, $\beta$=1.3) are
 considered to be representative of the average infrared spectrum used in 
 the model calculation.\\
\pn
 The spectral templates at FIR and submm wavelenghts have been
 normalized to the X--ray model spectra looking for correlations
 between the monochromatic luminosities at 30 keV and 100 \micron 
 for each class of \NH. We note that at 30 keV
 the source luminosity is not affected by absorption.
 The choice of this energy assures an unbiased estimate of the intrinsic 
 nuclear emission.\\
\pn
 It is then possible to integrate eq. (1) using the 
 luminosity function, redshift evolution and absorption
 distribution adopted in the XRB synthesis model and then
 calculate the AGN contribution to the CFIRB.
 The adopted values for the Hubble
 constant and the cosmological deceleration parameter are
 $H_0=50$ Km s\uno Mpc\uno and $q_0=0$.

\begin{figure}[t]
\rule{5cm}{0.2mm}\hfill\rule{5cm}{0.2mm}
\epsfig{figure=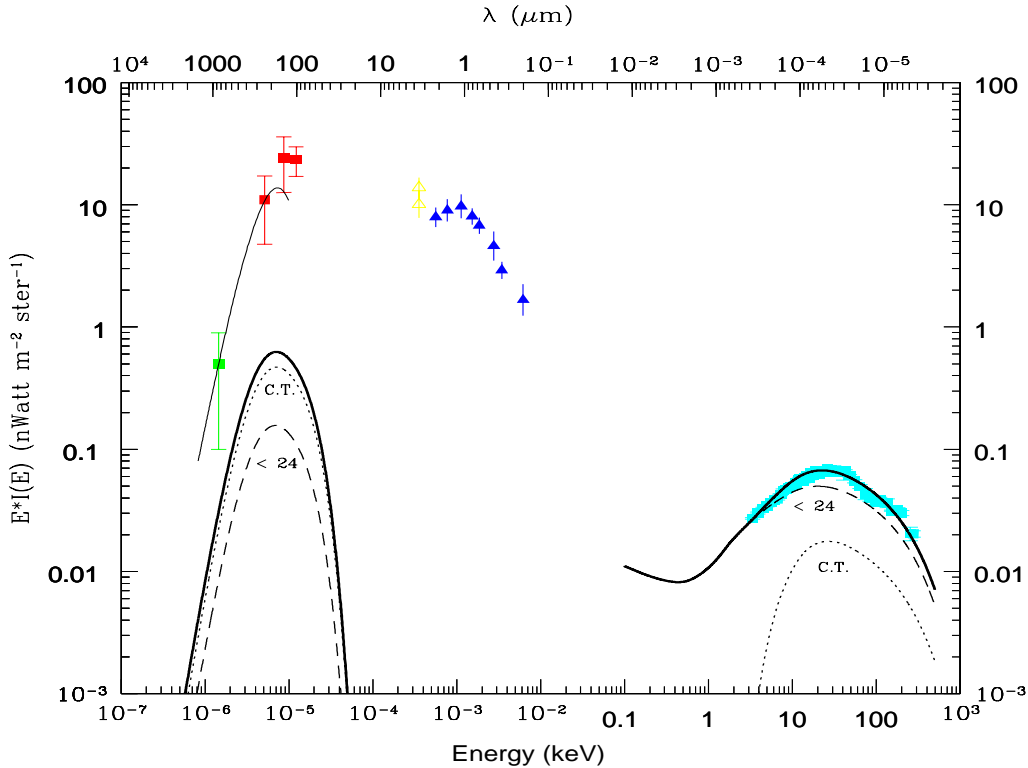,height=5.0in,width=6.0in}
\caption{The model predicted AGN contribution at FIR and X--ray
 wavelenghts (solid thick line). The dashed line represents the
 contribution of Compton--thin sources, while the dotted line that 
 of Compton--thick objects. 
 Data points are from a compilation of 
 measurements of the Extragalactic Background intensity from Fixsen \etal~1998
 (submm data and best--fit curve --- green point and black curve), 
 Lagache \etal~2000 (FIR data --- red points), Dwek \& Arendt~1998 
 (DIRBE Mid--IR data --- yellow points), Pozzetti \etal~1998 
 (optical and Near--IR data --- blue points) and Marshall \etal~1980
 (X--ray data --- cyan points)}
\rule{5cm}{0.2mm}\hfill\rule{5cm}{0.2mm}
\end{figure}

\section{Results}

 The model predictions are reported in Fig. 1, along with a
 compilation of recent measurements of the Extragalactic
 Background from FIR to X--rays.\\
 It is clear, from a visual inspection of the figure, that the 
 sources making most of the XRB do not significantly contribute 
 to the CFIRB (thick solid lines in Fig. 1).\\ 
 In order to test the sensitivity of the results to the 
 model parameters, we have run a number of different
 models by modifying the input parameters (dust temperature, slope
 and normalization of the L$_{IR}-L_X$ correlation) within 1$\sigma$ 
 from their best fit values.\\
 The results suggest that the AGN contribution is never dominant,
 being always in the range 3-15\% when compared to the 
 850 \micron energy density ($\nu I_{\nu}=0.5$ nW 
 m\due sr$^{-1}$, Fixsen \etal 1998), and of the order of a few percent in the
 100-200 \micron range (dashed region in Fig. 2). \\
 According to unified models for Active Galaxies, obscured AGN
 are characterized by a large amount of dust and gas able to 
 reprocess the primary radiation. It is not surprising that the 
 most important AGN contribution to the CFIRB is due to Compton--thick
 sources, with \NH in excess of $10^{24}$ cm\due (dotted line in Fig. 1).
 On the other hand, the bulk of the XRB is accounted for by Compton--thin
 AGN (\NH$<10^{24}$ cm\due, dashed line in Fig. 1).

\begin{figure}[t]
\rule{5cm}{0.2mm}\hfill\rule{5cm}{0.2mm}
\epsfig{figure=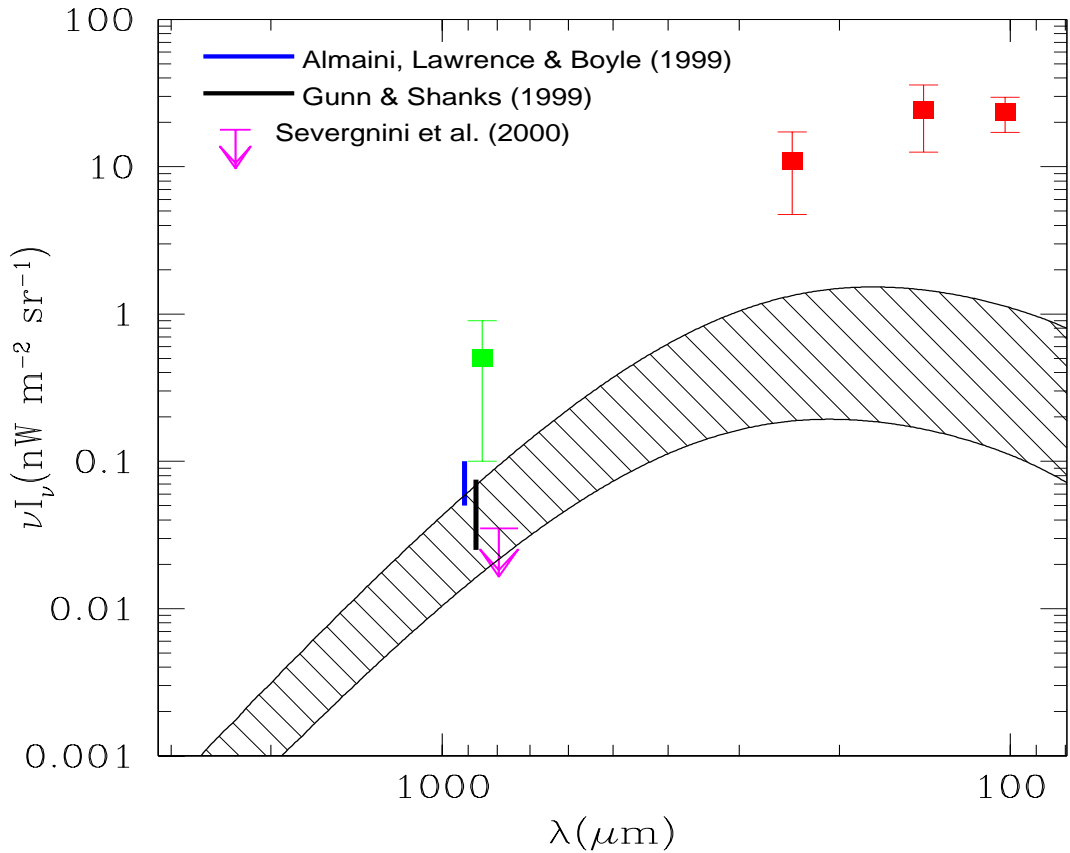,height=5.5in,width=6.5in}
\caption{A comparison of model predictions (shaded region)
 with the Almaini, Lawrence \& Boyle 1999 (cyan)
 and Gunn \& Shanks 1999 (yellow) results. 
 The upper limit estimated by Severgnini \etal~2000 combining deep 
 \textit{Chandra} and SCUBA observations is also reported.
 The data points are the same as in Fig. 1. The various model predictions are
 referred to 850 $\mu$m, but slightly shifted for clarity}  
\rule{5cm}{0.2mm}\hfill\rule{5cm}{0.2mm}
\end{figure}

\section{Discussion}

 The AGN contribution to the CFIRB has been calculated 
 by  Almaini, Lawrence \& Boyle (1999) and Gunn 
 \& Shanks (1999), following a similar approach, but 
 with different assumptions concerning the broad--band 
 AGN spectral energy distribution and their cosmic evolution.
 A good agreement is found between
 our results and their estimate (see Fig. 2): 
 the predicted AGN contribution at 850 \micron
 is in the range 3-20\% for all the models.\\
\pn
 Our results are also consistent with the observed
 (anti)--correlation between X--ray and FIR/submm 
 sources content at limiting fluxes where a large
 fraction of the backgrounds in the two bands is resolved
 (Fabian \etal~2000, Hornschmeier \etal~2000, Barger \etal~2001).
 Recent observational results are also reviewed
 by Lawrence (2001) and Hauser \& Dwek (2001): they both stress that 
 AGN contribute at most only $10-20\%$ to the CFIRB.
 An even more tight constraint ($<7\%$) has been obtained 
 by Severgnini \etal~(2000).\\ 

\pn
 As already pointed out in previous works, we note that
 our results should be considered as lower limits of the AGN
 contribution to the CFIRB.
\pn
 Indeed our approach is biased toward X--ray bright AGN:
 as a consequence, the observational L$_X$--L$_{IR}$ relation 
 obtained from an X--ray selected sample favours a low IR/X ratio.
 A possible important contribution from FIR bright, X--ray weak
 sources might have been underestimated or not be accounted at all.
\pn
 Moreover, according to our calculations, the most important
 contribution to the CFIRB comes from heavily obscured sources 
 (\NH$>10^{24}$ cm\due). The relative number of Compton--thick 
 AGN is only poorly constrained by XRB synthesis model
 which are not sensitive to their precise numerical fraction.
 It is thus possible to accomodate a larger 
 number of Compton--thick AGN without exceeding the XRB
 observational constraints and, at the same time, to 
 increase the contribution to the CFIRB. 
 Recent observational evidences do indeed indicate
 that the fraction of obscured objects in the local universe is
 higher than previously thought (Risaliti \etal~1999).
\pn 
 Finally, as already recognized by Almaini, Lawrence \& Boyle (1999) the 
 contribution to the CFIRB is strongly dependent from the AGN
 evolution at high redshift, adding further uncertainties to the model 
 predictions.\\
\pn
 A more detailed analysis of the parameter space, including an
 extension of model predictions to Mid--IR wavelenghts,
 is the subject of a paper in preparation (Brusa, Comastri \& Vignali 2001).\\

\section*{Acknowledgments}
The authors aknowledge partial support by ASI ARS/99/75 contract and
MURST Cofin-98-032 and Cofin-00-02-36 contracts.
This research has made use of the NASA/IPAC Extragalactic Database
(NED) which is operated by the Jet Propulsion Laboratory, Caltech,
under agreement with the National Aeronautics and Space Association.

\end{document}